\documentclass[11pt]{article}

\pdfoutput=1

\usepackage{fullpage}
\usepackage{amsmath}
\usepackage{amssymb}
\usepackage{amsthm}
\usepackage{mathrsfs}
\usepackage{todonotes}
\usepackage{xspace}
\usepackage{hyperref}
\usepackage{setspace}
\usepackage{epsfig}
\usepackage{graphicx}
\usepackage{chngcntr}
\usepackage{authblk}
\usepackage{pbox}
\usepackage{subfigure}

\usepackage{algorithm}
\usepackage[noend]{algpseudocode}   
\renewcommand\algorithmicdo{}	    
\algnewcommand\algorithmicendparfor{\textbf{end}}
\MakeRobust{\Call} 

      \algblock{ParFor}{EndParFor}
      \algblock{ForEach}{EndForEach}
      \algnewcommand\algorithmicparfor{\textbf{parallel for}}
      \algrenewtext{ParFor}[1]{\algorithmicparfor\ #1\ \algorithmicdo }
      \algrenewtext{EndParFor}{\algorithmicendparfor}
      \algnewcommand\algorithmicforeach{\textbf{for each}}
      \algrenewtext{ForEach}[1]{\algorithmicforeach\ #1\ \algorithmicdo }
      \algrenewtext{EndForEach}{\algorithmicendfor}

      \makeatletter
      \ifthenelse{\equal{\ALG@noend}{t}}%
	 {%
	 \algtext*{EndParFor}%
	 \algtext*{EndForEach}%
	 }{}%
      \makeatother

\newtheorem{theorem}{Theorem}

\newtheorem{lemma}[theorem]{Lemma}

\newcommand{\Ceil}[1]{\left\lceil #1 \right\rceil}

\newcommand{\Paren}[1]{\left ( #1 \right )}

\newcommand{\BigO}[1]{O\Paren{#1}}

\newcommand{\struct}{\textsc{parBucketHeap}\xspace}
\newcommand{\sssp}{\textsc{parDijsktra}\xspace}
\newcommand{\op}[1]{\textsc{Op}(#1)}
\newcommand{\res}[2]{Res_{#1}(#2)}

\allowdisplaybreaks

\title{A Parallel Priority Queue with Fast Updates for GPU Architectures}
\author[1]{Kyle Berney}
\author[2]{John Iacono}
\author[2]{Ben Karsin}
\author[1]{Nodari Sitchinava}

\affil[1]{University of Hawaii at Manoa}
\affil[2]{Universit\'e Libre de Bruxelles}

\date{}

\begin{document}
\maketitle

\begin{abstract}
The single-source shortest path (SSSP) problem is a well-studied problem that is used in many applications.
In the parallel setting, a work-efficient algorithm that additionally attains $o(n)$ parallel depth has been elusive.
Alternatively, various approaches have been developed that take advantage of specific properties of a particular class of graphs.
On a graphics processing unit (GPU), the current state-of-the-art SSSP algorithms are implementations of the Delta-stepping algorithm, which does not perform well for graphs with large diameters.
The main contribution of this work is to provide an algorithm designed for GPUs that runs efficiently for such graphs.

We present the parallel bucket heap, a parallel cache-efficient data structure adapted for modern GPU architectures that supports standard priority queue operations, as well as bulk update.
We analyze the structure in several well-known computational models and show that it provides both optimal parallelism and is cache-efficient.
We implement the parallel bucket heap and use it in a parallel variant of Dijkstra's algorithm to solve the SSSP problem.
Experimental results indicate that, for sufficiently large, dense graphs with high diameter, we outperform the current state-of-the-art SSSP implementations on an Nvidia RTX 2080 Ti and Quadro M4000 by up to a factor of 2.8 and 5.4, respectively.
\end{abstract}

\section{Introduction}
In the past decade, graphics processing units (GPUs) have emerged as an effective hardware architecture for solving computationally intensive problems.
Modern GPUs provide thousands of physical cores, low latency memory units, and fast context switching between threads.
The high computational throughput of GPUs have made it one of the most frequently used hardware systems for large compute clusters and supercomputers.

However, the complexity of the GPU architecture poses a challenge for both theoretical (design and analysis of algorithms) and experimental research (providing fast implementations).
A typical GPU-efficient algorithm requires a high degree of parallelism, while still exhibiting optimal memory access patterns to various memory units (e.g., global memory, shared memory).
Additionally, many interdependent factors must be considered, such as: the hierarchical organization of threads, hardware resource limitations (e.g., amount of register space available), synchronization between various levels of the thread hierarchy, and the maximum number of active threads resident on the GPU (called \emph{occupancy}).
As a result, it can be especially challenging to provide efficient and fast implementations of GPU algorithms for irregular computations, such as various graph algorithms.

Single-source shortest paths (SSSP) is a fundamental graph problem that has applications in many domains.
Let $G = (V, E)$ be a directed graph consisting of $|V| = n$ vertices and $|E| = m$ non-negative weighted edges.
Given a source vertex $v \in V$, the SSSP problem asks to find the minimum weight path from $v$ to all other reachable vertices $u \in V$.
In the sequential setting, the two classical solutions are Dijkstra's algorithm~\cite{dijkstra} and Bellman-Ford~\cite{bellman,ford}.
Both take an iterative approach, where vertices are labeled with a tentative distance from the source vertex (initially set to $-\infty$) and are iteratively updated throughout execution.
The difference in the algorithms comes in the order in which edges are processed.
In each iteration of Dijkstra's algorithm, the outgoing edges of the minimum distance vertex, which has not been visited yet, are processed.
Hence, Dijkstra's algorithm (using a Fibonacci heap) uses $O(m + n\log{n})$ total operations.
In contrast, Bellman-Ford performs $O(nm)$ total operations as all edges are processed in each iteration.
Consequently, Bellman-Ford is easily parallelizable and is able to additionally compute shortest paths on graphs with negative edge weights.

In the parallel setting, a work-efficient algorithm with fast parallel runtime (e.g., $o(n)$) is yet to be developed.
Instead, numerous parallel algorithms have been proposed that sacrifice work-efficiency for increased parallelism.
Typically, these algorithms are designed to take advantage of specific properties of a particular class of graphs (e.g., random graphs, planar graphs, etc.).

The current state-of-the-art implementations of SSSP on GPUs are variations of the Delta-stepping algorithm of Meyer et al.~\cite{delta-stepping}, which has been proven to run well on graphs with random edge weights, ``small'' maximum vertex degrees, and ``small'' maximum shortest path lengths.
In this work, we focus on graphs with sufficiently large diameter and degrees.
We present a parallelization of the cache-oblivious bucket heap of Brodal et al.~\cite{brodal-bucketheap} and buffer heap of Chowdhury and Ramachandran~\cite{chowdhury-buffer} and adopt it for GPU architectures.
Using the resulting heap, we implement a parallel variant of Dijkstra's algorithm and compare it to the state-of-the-art GPU SSSP implementations.
Throughout this paper, we assume the reader is familiar with the GPU architecture, in particular, the thread and memory hierarchy of GPUs (see~\cite{cuda-best, cuda-c} for more information).

\subsection{SSSP}
\label{sec:sssp}
In the parallel setting, the SSSP problem suffers from the \emph{transitive closure bottleneck}~\cite{karp}.  
Thus, finding an algorithm that is work-efficient (i.e., the same work complexity as Dijkstra's algorithm) with $o(n)$ runtime on an arbitrary graph remains an important open problem.
As a result, many alternative parallel SSSP algorithms have been proposed for specific classes of graphs.

For planar graphs with integer edge weights between $0$ and $k$, Klein and Subramanian~\cite{klein} solve SSSP in $O(\textnormal{polylog } n \log{k})$ parallel time using $n$ processors. 
Subramanian et al.~\cite{subramanian} show that planar layered directed graphs can be decomposed using one-way separators that results in an SSSP algorithm with $O(\log^3{n})$ parallel runtime using $n$ processors. 
Tr{\"{a}}ff and Zaroliagis~\cite{traff} use a region decomposition of a planar directed graph and show that for $0 < \epsilon < \frac{1}{2}$, their SSSP algorithm has $O((n^{2\epsilon} + n^{1-\epsilon})\log{n})$ depth and $O(n^{1+\epsilon})$ work. 
Atallah et al.~\cite{atallah} present a $O(\log^2{n})$ depth and $O(n\log{n})$ work SSSP algorithm for planar layered directed graphs. 

Chaudhuri and Zaroliagis~\cite{chaudhuri} consider directed graphs with constant treewidth (a measure of how ``close'' the graph is to a tree) and use a $O(\log^2{n})$ depth and $O(n)$ work preprocessing stage that allows the computation of the SSSP with path length $\ell$ in $O(\alpha(n) \log{n})$ depth and $O(\ell + \alpha(n) \log{n})$ work \footnote{$\alpha(n)$ is the inverse Ackermann function}. 

For directed graphs with negative integer weights lower bounded by some integer $-k$, Cao et al.~\cite{cao} present a parallelization of Goldberg's algorithm~\cite{goldberg} that solves SSSP in $n^{5/4+o(1)}\log{k}$ depth and $\tilde{O}(m\sqrt{n}\log{k})$ work, with high probability
\footnote{$\tilde{O}$ hides polylogarithmic factors that may be present in the standard $O$ notation}.

Crauser et al.~\cite{crauser} divide Dijkstra's algorithm into phases and show that on random graphs with random edge weights, the algorithm has $O(n^{\frac{1}{3}} \log{n})$ depth and $O(m + n\log{n})$ work with high probability\footnote{For some constant $c > 0$, the probability is at least $1 - n^{-c}$} on a CRCW PRAM. 
Meyer et al.~\cite{delta-stepping} introduce the Delta-stepping algorithm, where in each iteration, the outgoing edges of vertices within a distance interval of width $\Delta$ are processed.
For an arbitrary graph with random edge weights, maximum degree $d$, maximum shortest path distance $L$, and $\Delta = \Theta\left(\frac{1}{d}\right)$; the Delta-stepping algorithm has a parallel depth of $O(dL \log{n} + \log^2{n})$ and total work of $O(n + m + dL\log{n})$ on average. 

For undirected graphs, Spencer and Shi~\cite{shi} first compute the $k$ nearest neighbors of every vertex in $O(\log{n} \log{k})$ depth and $O(nk^2\log{n}\log{k} + m)$ work and use this information to solve SSSP in $O(\frac{n}{k}\log{n})$ depth and $O((m + nk)\log{n})$ work. 
Blelloch et al.~\cite{radius-stepping} combine the approaches of Spencer and Shi~\cite{shi} and Meyer et al.~\cite{delta-stepping} for undirected $(k, \rho)$-graphs, which are graphs where every vertex can reach its $\rho$ closest neighbors in $k$ or fewer edges traversed. 
Their SSSP algorithm has $O(\frac{kn}{\rho} \log{n}\log{\rho L})$ depth and $O(km\log{m})$ work.
The authors additionally provide a preprocessing stage that transforms any undirected graph into a $(1, \rho)$-graph with at most $n\rho$ additional edges in $O(\rho \log{\rho})$ depth and $O(m \log{n} + n\rho ^ 2)$ work. 

Considering approximate solutions on undirected graphs, Cohen~\cite{cohen} defines a $(d, \epsilon)$-hop set of a graph, which augments the graph with new edges such that the shortest path, consisting of at most $d$ edges, in the new graph has a distance within $(1 + \epsilon)$ of the shortest path in the original graph.
Let $\epsilon_0 > 0$ be a fixed constant, Cohen presents a randomized algorithm that constructs 
a $(O(\textnormal{polylog } n), O(1 / \textnormal{polylog } n))$-hop set with $O(n^{1+\epsilon_0})$ edges and uses it to solve the approximate shortest path problem from $s$ different sources using $O(mn^{\epsilon_0} + s(m + n^{1 + \epsilon_0}))$ work and polylogarithmic parallel runtime. 
Building on this work, Elkin and Neiman~\cite{elkin16} devise an alternate randomized construction of a $(O(1), O(1 / \textnormal{polylog } n))$-hop set with $O(n^{1+\epsilon_0} \log{n})$ edges that is used to improve the parallel runtime of the approximate shortest path problem. 
And in 2019, Elkin and Neiman~\cite{elkin19} present a randomized construction of a $(O(1), O(1 / \textnormal{polylog } n))$-hop set with $O(n^{1+\epsilon_0} \log^*{n})$ edges.

Within the context of GPUs, Harish et al.~\cite{harish} showed that a GPU implementation of Bellman-Ford outperforms a sequential CPU approach.
More recently in 2014, Davidson et al.~\cite{owens-sssp} experimentally evaluated several GPU implementations of SSSP.
Notably, a variation of Bellman-Ford, called Workfront Sweep, and an implementation of the Delta-stepping algorithm of Meyer et al.~\cite{delta-stepping}, called Near-Far.
Workfront Sweep uses a heuristic that seeks to reduce the amount of work performed in each iteration of Bellman-Ford by only processing edges outgoing from vertices whose tentative distances were updated in the previous iteration.
Let $W$ be the average weight of edges in the graph, $d'$ be the average degree in the graph, and $w$ be the number of threads in a warp of a GPU.
The Near-Far implementation uses $\Delta = \frac{W \cdot w}{d'}$ and only two buckets (the ``near'' and ``far'' buckets).
Experiments were conducted on 8 different graphs and results showed that the Near-Far implementation provides performance gains on 6 graphs with low diameter and degree (5 out of the 6 graphs have random edge weights).
Building on the Workfont Sweep approach, Busato and Bombieri~\cite{busato} provide a variation of Bellman-Ford that additionally classifies edges based on the operations needed to process the edge (e.g., whether an atomic operation is needed).
In 2016, Wang et al.~\cite{gunrock} introduced the Gunrock library that contains an implementation of the Near-Far approach and showed that on graphs with low degree and random edge weights between 1 and 64, their implementation outperforms both CPU and GPU libraries for SSSP.
Lastly in 2021, Wang et al.~\cite{adds} improved on the Near-Far approach by using a heuristic that periodically changes the $\Delta$ value, adding a dynamic memory allocator to allow for the use of multiple buckets, and using a designated group of threads to manage the coordination of work.

\subsection{Priority Queues}
\label{sec:prioqueue}
Dijkstra's algorithm for solving SSSP relies on an efficient priority queue to find the vertex with the minimum tentative distance that has not been visited yet.
Formally, the priority queue ADT is defined over a collection of elements, $Q$, where each element $e$ consists of a value and priority, i.e., $e \in Q = (val, p)$.
For each element $e$, we define $e.val$ and $e.p$ to be the value and priority, respectively.
The ADT supports the following operations:
\begin{itemize}
\item \textsc{extractMin} removes and returns the element in $Q$ with the smallest priority, i.e., $e \in Q$ such that $e.p = \min\limits_{e_i \in Q} e_i.p$

\item $\textsc{update}(e)$ adds new element $e = (val,p)$ to $Q$, and if there exists an $e' = (val, p')$ with the same value in $Q$, then remove $e'$ (so that it is replaced by $e$)\footnote{In this work, we assume that updates only decrease priority. However, this assumption can be removed by adding a timestamp to each element when it is inserted into the structure; and when deleting duplicate entries, the most recent timestamp is kept.}

\item $\textsc{delete}(e)$ removes $e$ from $Q$ if it is contained in $Q$
\end{itemize}

There are many data structures defined that implement the priority queue ADT, including various types of heaps~\cite{brodal-bucketheap,brodal-parheap,chowdhury-buffer,CLRS,dietz,fredman-pairing,fredman-fib}.
We note that many other data structures, including binary search trees or sorted arrays, can be used as priority queues, though they provide additional functionality and are therefore not as efficient when performing only the above operations.
Though not considered part of the standard priority queue ADT, in this work we additionally consider the $\textsc{BulkUpdate}(U)$ operation that, given a set of elements $U$, performs $\textsc{Update}(e)$ for each $e \in U$.
In each iteration of Dijkstra's algorithm, all outgoing edges of the current minimum distance vertex are processed, resulting in a set of new (shorter) tentative distances.
The $\textsc{BulkUpdate}(U)$ operation is used to update the tentative distances of these vertices in a single efficient operation, thereby allowing efficient processing of graphs with large degrees.

Several fundamental priority queue data structures provide tradeoffs between simplicity and performance (e.g., binary heaps, Fibonacci heaps~\cite{fredman-fib}, pairing heaps~\cite{fredman-pairing}, etc.).
However, these heaps are inherently sequential and operations on heaps cannot be easily parallelized.  Thus, several priority queues have been developed to expose parallelism~\cite{brodal-parheap,dietz,topk,hunt,randomprio}.
Brodal et al.~\cite{brodal-parheap} presents a structure that performs all standard priority queue operations in constant time and logarithmic work in the PRAM model.
H{\"{u}}bschle{-}Schneider et al.~\cite{topk} design a randomized parallel priority queue that supports \textsc{BulkUpdate} on up to $d$ elements in $O(1 + \log{d})$ parallel time, in a parallel distributed memory model (i.e., processors communicate over an interconnection network).

To our knowledge, all existing parallel priority queue data structures are not cache-efficient, and as such may not perform well on GPUs or other parallel systems that rely on locality of reference to achieve peak performance.
While not inherently parallel, in the context of the External Memory or cache-oblivious models, the cache-oblivious bucket heap~\cite{brodal-bucketheap} and buffer heap~\cite{chowdhury-buffer} structures achieve sub-constant amortized time operations when the block size, $B$, is sufficiently large (see Section~\ref{sec:models} for a definition).
Since there are no parallel, cache-efficient priority queue structures, few works have considered using priority queues on GPUs: 
He et al.~\cite{he} present a priority queue that achieves a speed up factor of 30 over sequential execution; and Baudis et al.~\cite{GPU-circle-buffer} demonstrate that for small queues of up to 500 items, simple circular buffers outperform tree-based queues when evaluated on discrete event simulation and A$^{*}$ search.

\begin{table*}[t]
\footnotesize
\label{table:priority-queue}
\begin{tabular}{|c|c|c|c|c|c|}
\hline
& \multicolumn{2}{|c|}{Time} & \multicolumn{2}{|c|}{I/O Complexity} & \multicolumn{1}{|c|}{Total I/Os} \\
Data structure & \textsc{ExtractMin} & \textsc{BulkUpdate} & \textsc{ExtractMin} & \textsc{BulkUpdate} & $\frac{n}{d}$ \textsc{BulkUpdate}s\\
\hline
\pbox[c]{20cm}{\strut Seq. Bucket \\ Heap~\cite{brodal-bucketheap}\strut} & $O(\log{n})$ & $O(d\log{n})$ & \strut $O\left(\frac{1}{B} \log{\frac{n}{B}}\right)$\strut  & $O\left(\frac{d}{B} \log{\frac{n}{B}}\right)$ & $O\left(\frac{n}{B} \log{\frac{n}{B}}\right)$\\
\hline
\pbox[c]{20cm}{\strut Parallel Prio. \\ Queue~\cite{brodal-parheap}\strut} & $O(1)$ & $O(1)$ & $O(1)$ & $O(1)$ & $O(n\log{\frac{n}{d}})$\\
\hline
\pbox[c]{20cm}{\strut Bulk Parallel \\ Prio. Queue~\cite{topk}\strut} & $O(1 + \log{d})$ & $O(1 + \log{d})$ & - & - & - \\
\hline
This work & $O(1 + \log{d})$ & $O(1 + \log{d})$ & \strut $O\left(\frac{\log{(n/d)}}{pB} + \frac{1}{B}\right)$ \strut  & $O\left(\frac{d\log{(n/d)}}{pB} + \frac{d}{B}\right)$ & $O\left(\frac{n}{B}\log{\frac{n}{d}}\right)$\\
\hline
\end{tabular}
\caption{
Comparison of priority queue operations in different sequential and parallel models: $n$ is the number of input elements, $d \leq n$ is then maximum number of elements supported by \textsc{BulkUpdate}, $p$ is the number of processors, and $B$ is the width of data transfers to external memory. 
The right-most column shows the total number of I/Os when performing $\frac{n}{d}$ \textsc{BulkUpdate} operations, each consisting of $d$ updates.
(We note that the Bulk Parallel Priority Queue~\cite{topk} is designed in a parallel distributed memory model, hence, does not include I/Os to external memory.)
}
\end{table*}

\subsection{Models of Computation}
\label{sec:models}
While several algorithmic models for GPUs have been introduced~\cite{hong,agpu,tmm,nakano,hmm}, none of them has yet been widely adopted.
One reason for this is the complexity of the GPU architecture, which leads to an abundance of interdependent factors that can be justified and considered in these models.
However, common performance metrics used in these models are: 
the maximum number of operations executed by any single thread, the total number of operations executed across all threads, and the number of parallel \emph{coalesced} accesses into global memory.
Typically these metrics are then considered with other factors such as number of cores per SM, latency, or bandwidth.

Alternatively, research in GPU algorithms has mostly focused on optimizing a small number of specific performance metrics (e.g., global memory accesses, shared memory accesses, parallel time, or total work). 
Often, well-known parallel algorithmic models are used to perform analysis of these metrics.
Specifically, the Parallel Random Access Machine (PRAM) model~\cite{pram} is used to measure parallelism~\cite{berneysearch, cederman, cong14, mergepath, kimbook} and the Parallel External Memory (PEM) model~\cite{pem} is used to measure the number of parallel coalesced accesses to global memory~\cite{berneysearch, casanova17, kimbook}.
We highlight the fact that adopting established parallel models for developing and analyzing aspects of GPU algorithms has the additional advantage of leveraging the vast literature of algorithms and techniques that have been developed in these models.
In this work, we adopt this approach and use both the PRAM and PEM models to analyze the asymptotic performance of our algorithms.

Given an input size of $n$ elements, the PRAM model defines two performance metrics: \textit{work}, denoted $W(n)$, is the total number of operations performed by all processors; and \textit{depth}, denoted $D(n)$, is the maximum number of operations performed by any single processor if the algorithm is executed using an infinite number of processors.
Using Brent's Scheduling Principle~\cite{brent}, the runtime of an algorithm on $p$ processors can be computed as $T(n, p) = \BigO{\frac{W(n)}{p} + D(n)}$.
As the behavior of concurrent writes is undefined on GPUs, we consider the CREW (concurrent read, exclusive write) PRAM model that allows the concurrent reading but disallows the concurrent writing to the same memory location by different processors.

The PEM model is a parallel extension of the sequential External Memory (EM) model~\cite{em}.
In the EM model, a single processor contains an internal memory space of size $M$ and data is transferred between external and internal memory in blocks of contiguous memory of size $B$.
In order to process any data, that data element must reside in internal memory.
The performance metric in the EM model, called \textit{I/O complexity}, is the total number of such block transfers.
In the PEM model, $p$ processors each with an internal memory space of size $M$ are connected to a shared external memory space.
Each processor still transfers blocks of contiguous data between external memory and its own private internal memory.
The \emph{parallel I/O complexity} 
is defined as the maximum number of block transfers by any single processor.
On a GPU, $w$ contiguous global memory locations are able to be transferred to a group of $w$ threads, called a \textit{warp}. 
It has been observed in previous work~\cite{berneysearch,casanova17,kimbook,agpu,tmm,nakano,hmm} that this behavior is equivalent to block access (e.g., $B = w$ in the PEM model).
Thus, we can utilize the PEM model to analyze the number of parallel coalesced global memory accesses on a GPU.

\subsection{Our Contributions}
\label{sec:contributions}
We present the parallel bucket heap, denoted \struct, an I/O efficient parallel priority queue designed for GPU architectures supporting the \textsc{BulkUpdate} operation.
Using the \struct, the number of I/Os performed for a sequence of \textsc{BulkUpdate} operations is significantly reduced compared to the current best data structures (see Table~\ref{table:priority-queue}).
The \textsc{BulkUpdate} operation is particularly useful when the \struct is used to solve the SSSP problem using Dijkstra's algorithm, as batches of update operations are performed when the vertex being processed has multiple outgoing edges.

We use the \struct to implement a parallel version of Dijkstra's algorithm, denoted \sssp.
Both \struct and \sssp are implemented using CUDA C/C++~\cite{cuda}.
Experiments are conducted on 2 NVIDIA GPUs: an RTX 2080 Ti and a Quadro M4000.
We compare the performance of \sssp to the current state-of-the-art SSSP GPU implementations: Gunrock~\cite{gunrock} and Asynchronous Dynamic Delta-Stepping (ADDS)~\cite{adds}.
Our results show that for sufficiently dense graphs with large diameter ($n = 30,000$ vertices and diameter $n-1$), \sssp using the \struct has a peak speed up of 2.8 and 12 over Gunrock and ADDS, respectively, on the RTX 2080 Ti; and a peak speed up of 5.4 over Gunrock on the Quadro M4000 (ADDS does not support the Quadro M4000).

The paper is organized as follows: in Section~\ref{sec:bucketheap} we provide an overview of the sequential bucket heap and present our \struct data structure; in Section~\ref{sec:analysis} we analyze the \struct in the CREW PRAM and PEM models; in Section~\ref{sec:experiments} we provide implementation details and experimental results; and lastly, in Section~\ref{sec:conclusion} we conclude with a brief summary.

\section{Bucket Heap}
\label{sec:bucketheap}
\begin{figure*}[t]
  \centering
  \includegraphics[width=0.65\textwidth]{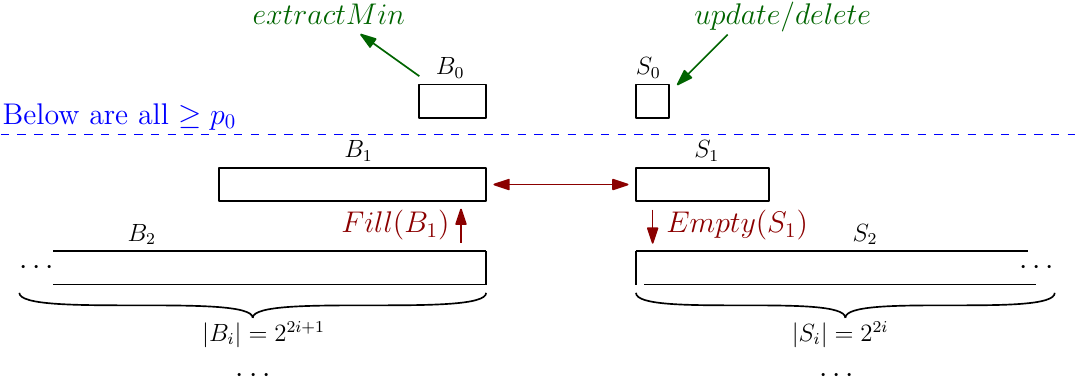}
  \caption{Illustration of the sequential bucket heap structure of Brodal et al.~\cite{brodal-bucketheap}. \textsc{update}s and \textsc{delete}s are inserted into $S_0$, while $\textsc{extractMin}$s are removed from $B_0$.  $\textsc{Empty}(S_i)$ empties $S_i$ into $S_{i+1}$, $\textsc{Fill}(B_i)$ fills $B_i$ from $B_{i+1}$, while $p_i$ is maintained at each level to ensure the heap property between levels.}
  \label{fig:bucketheap}
\end{figure*}

The parallel bucket heap is a parallelization of the cache-oblivious bucket heap of Brodal et al.~\cite{brodal-bucketheap} and buffer heap of Chowdhury and Ramachandran~\cite{chowdhury-buffer}.
In this paper, we follow the naming conventions and presentation of the bucket heap, thus, we first provide a general overview of the sequential bucket heap.

\subsection{Sequential Bucket Heap}
\label{sec:seqbucketheap}
The bucket heap is a hierarchical data structure, where each level consists of a \emph{bucket} and a \emph{signal buffer}.
We note that elements are always stored in sorted order by value (not priority) in each bucket and signal buffer.
Figure~\ref{fig:bucketheap} illustrates the sequential bucket heap structure and shows the relationship between elements stored at each level.
Elements inserted into the bucket heap (via \textsc{Update} operations) are moved into the top level's signal buffer; and elements removed from the bucket heap (via \textsc{ExtractMin} operations) are taken from the top level's bucket.
For any given level, if its signal buffer becomes sufficiently full (e.g., at least half full), then it is emptied into the bucket on the same level and overflow elements are merged into the next (lower) level's signal buffer.
And if the bucket becomes too empty (e.g., at least half empty), then it is filled with the smallest priority elements from the next (lower) level's bucket and signal buffer.

Formally, a bucket heap storing $n$ elements has $q = \Ceil{\log_4{n}} + 1$ levels, where for each level $i \in \{0, 1, \ldots, q-1\}$, the maximum capacity of the $i$-th level's bucket, denoted $B_i$, and signal buffer, denoted $S_i$, is $2^{2i+1}$ and $2^{2i}$, respectively.
The bucket heap maintains the invariant that for all $j > i$, all elements in $B_i$ have a smaller (or equal) priority than all elements in $B_{j}$.
This ensures that if $B_0$ is non-empty and $S_0$ is empty, then $B_0$ will contain the minimum priority element in the structure.
Therefore, if this condition is satisfied, $\textsc{ExtractMin}$ can simply remove and return the minimum priority element in $B_0$.
Futhermore, as long as $S_0$ is kept non-full, $\textsc{Update}(e)$ can simply insert $e$ into $S_0$.
$\textsc{Delete}$ operates similar to $\textsc{Update}$ using an element with a special priority value, \textsc{DEL}, that moves down the structure, removing elements with matching key values.

The bucket heap transfers elements between levels via the $\textsc{Empty}$ and $\textsc{Fill}$ operations.
The $\textsc{Empty}(S_i)$ operates as follows: 
(1) scan $B_i$ to find the element with maximum priority, denoted $p_i$; 
(2) merge elements in $S_i$ with $B_i$; 
(3) for any elements with duplicate values, remove those with larger priority; and 
(4) all elements $e$ such that $e.p > p_i$ are merged into $S_{i+1}$.
If the resulting number of elements in $B_i$ is too full (i.e., there are too many elements with $e.p \leq p_i$) then $p_i$ is updated so that $|B_i| = 2^{2i+1}$ and the elements with priorities larger than $p_i$ are merged into $S_{i+1}$.
Updating the priority of an existing element is accomplished when elements with duplicate values are found and the element with larger priority is removed (elements with special delete priorities, \textsc{DEL}, are also applied this way).
Since lists are stored sorted by value, elements with duplicate values are stored next to each other and can be removed with a scan.
If the number of elements in a bucket $B_i$ fall under the minimum size (e.g., half full), $\textsc{Fill}(B_i)$ is called, which empties $S_i$ into $B_i$ and fills any remaining space in $B_i$ with elements from level $(i+1)$.
This is accomplished by:
(1) calling $\textsc{Empty}(S_i)$; 
(2) if $B_i$ is non-full and $S_{i+1}$ is non-empty, then $\textsc{Empty}(S_{i+1})$ is called; and
(3) if $B_i$ is non-full, then $B_i$ is filled with the smallest priority elements in $B_{i+1}$.
All of these operations are performed via scans of contiguous arrays, 
leading to $\BigO{\frac{1}{B}\log{\frac{n}{B}}}$ amortized I/O complexity of the $\textsc{ExtractMin}$, $\textsc{Update}$, and $\textsc{Delete}$ operations of the sequential bucket heap.

\subsection{Parallel Bucket Heap}
\label{sec:parbucketheap}
We parallelize the sequential bucket heap by using parallel variants of \textsc{Empty} and \textsc{Fill}; and allowing non-adjacent levels to execute in parallel.
Additionally, we increase the maximum capacity of every bucket and signal buffer by a factor of $d$, hence, $|B_i| = d \cdot 2^{2i+1}$ and $|S_i| = d \cdot 2^{2i}$.
When $U$ is sorted by value, a $\textsc{BulkUpdate}(U)$ of up to $d$ elements can be efficiently performed by simply inserting all updates into $S_0$.
By increasing the capacity of all buckets and signal buffers, we decrease the total number of levels of the bucket heap to $q = \Ceil{\log_4{\frac{n}{d}}} + 1$.

\begin{algorithm*}
\footnotesize
\begin{algorithmic}[1]
\Statex \textbf{Precondition:} if $i < \ell$, then $|B_{i}| \geq d \cdot 2^{2i}$
\Statex \textbf{Precondition:} $|S_{i}| \leq d \cdot 2^{2i}$
\Statex \textbf{Precondition:} if $i+1 < \ell$, then $|B_{i+1}| \ge d \cdot 2^{2(i+1)} + d \cdot 2^{2i}$
\Statex \textbf{Precondition:} $|S_{i+1}| \leq d \cdot 2^{2(i+1)} - d \cdot 2^{2i}$
\Statex \textbf{Postcondition:} if $i < \ell$, $|B_i| = d \cdot 2^{2i+1}$
\Statex \textbf{Postcondition:} $|S_i| = 0$

\If{$|S_i| > 0$}
\Comment{Empty $S_i$ if needed}
  \State $B_i \gets$ \Call{Merge}{$S_i$, $B_i$}; $S_i = \emptyset$
  \State $B_i \gets$ \Call{DeleteDuplicates}{$B_i$}
  \State $num \gets |\{e: e \in B_i \mbox{ and } e.p \leq p_i\}|$ \Comment{Count elements with small priority}
  \If{$num > 2^{2i+1}$} \Comment{Update $p_i$ if needed}
    \State $p_i \gets \Call{Select}{B_i, 2^{2i+1}}$
  \EndIf
  \State $B'_{i} = \{e: e \in B_i \mbox{ and } e.p > p_i\}$ \Comment{Move large priority elements to $S_{i+1}$}
  \State $B_{i} = \{e: e \in B_i \mbox{ and } e.p \le p_i\}$
  \State $S_{i+1} \gets \Call{Merge}{S_{i+1}, B'_{i}}$
\EndIf
\Statex
\If{$|B_i| < 2^{2i+1}$ and $i$ is not largest non-empty level}
\Comment{Fill $B_i$ if needed}
  \State $B_{i+1} \gets \Call{Merge}{B_{i+1}, S_{i+1}}$; $S_{i+1} \gets \emptyset$
  \State $B_{i+1} \gets \Call{DeleteDuplicates}{B_{i+1}}$
  \State $p_i \gets \Call{Select}{B_{i+1}, 2^{2i+1}{-}|B_i|}$ 
  \State $B'_{i+1} \gets \{e : e \in B_{i+1} \mbox{ and } e.p \le p_i\}$ \Comment{Pull elements up to fill $B_i$}
  \State $B_{i+1} \gets \{e : e \in B_{i+1} \mbox{ and } e.p > p_i\}$
  \State $B_i \gets$ \Call{Merge}{$B_i, B'_{i+1}$}
\State $S_{i+1} \gets \{e :e \in B_{i+1} \mbox{ and } e.p > p_{i+1} \}$ \Comment{Move large priority elements back to $S_{i+1}$}
\EndIf
\end{algorithmic}
\caption{{\sc Resolve}$(i)$}
\label{alg:resolve}
\end{algorithm*}

For ease of exposition, we combine the $\textsc{Empty}$ and $\textsc{Fill}$ operations into a single operation, $\textsc{Resolve}$ (Algorithm~\ref{alg:resolve}).
Let $\ell$ be the maximum non-empty level of the parallel bucket heap.
The $\textsc{Resolve}(i)$ operation empties $S_i$ and fills $B_i$, leaving $S_i$ empty and $B_i$ full (unless $i=\ell$).
Our description of the $\textsc{Resolve}$ operation in Algorithm~\ref{alg:resolve} is high-level and the subroutines $\textsc{Merge}$, $\textsc{DeleteDuplicates}$, and $\textsc{Select}$ can be implemented in different ways, depending on the desired level of parallelism, which we discuss in our analysis in the subsequent sections.

\subsubsection{Parallel Execution Sequence}
\label{sec:executionseq}
Consider a series of $N$ \emph{operations}, defined as $\textsc{Op}_{1}, \textsc{Op}_{2}, \ldots, \textsc{Op}_{N}$, where each operation is $\textsc{ExtractMin}$, $\textsc{Update}$, $\textsc{Delete}$, or $\textsc{BulkUpdate}$.
In the sequential setting, the bucket heap empties signal buffers and fills buckets as needed.
While in the parallel setting, we can proactively perform \textsc{Resolve} on different levels of the bucket heap in parallel.
Let $\res{i}{k}$ be the $k$-th execution of $\textsc{Resolve}(i)$ during the series of operations.

\begin{figure*}[tb]
  \centering
  \includegraphics[width=0.95\textwidth]{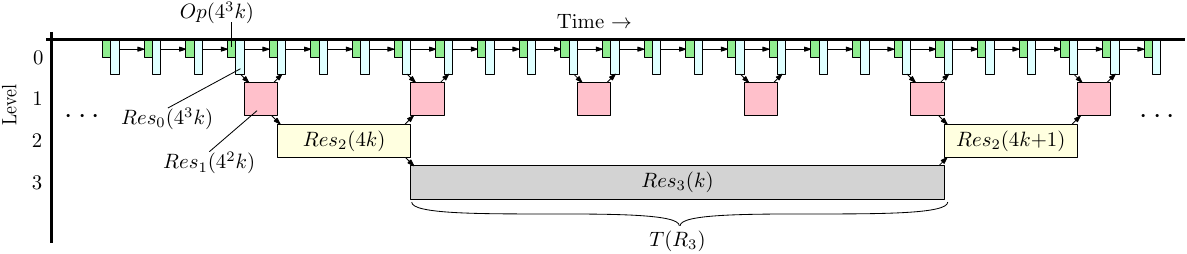}
  \caption{Illustration of the dependencies when performing a series of operations. Small green boxes represent operations ($\textsc{extractMin}$, $\textsc{update}$, $\textsc{delete}$, or $\textsc{bulkUpdate}$) and the remaining boxes represent $\textsc{Resolve}$s. Dependencies are shown with arrows between boxes. Since resolving larger levels takes more time, we represent them by wider boxes which are scaled to show that, if $\textsc{Resolve}(0)$ takes $d$ time then $\textsc{Resolve}(i)$ can take $d \cdot 2^{2i}$ time without delaying any operations (which occur every $5d$ parallel memory accesses).}
  \label{fig:resolution-schedule}
\end{figure*}

We define $A \rightarrow B$ to denote that task $B$ depends on task $A$ being completed in order for the preconditions of task $B$ to be satisfied.
Intuitively, we view the execution of \textsc{Resolve}s as a directed acyclic graph (DAG) where each vertex represents a \textsc{Resolve} operation and each edge is a dependency.
Figure~\ref{fig:resolution-schedule} illustrates this DAG, where green boxes represent operations, other color boxes represent \textsc{Resolve}s, and the width of each box is the amount of time needed to perform it.

\begin{theorem}
\label{thm:4-resolves}
Let $i > 0$ and $k \geq 1$.
If, for every level $i$, we perform $\textsc{Resolve}(i)$ after every fourth $\textsc{Resolve}(i-1)$, then all preconditions are always satisfied, i.e., $$\res{i-1}{4k} \rightarrow \res{i}{k}$$
\end{theorem}
\begin{proof}
After $\textsc{Resolve}(i)$ completes, $|S_i| = 0$ and $|B_i| = d \cdot 2^{2i + 1}$.
Each call to $\textsc{Resolve}(i - 1)$ adds at most $d \cdot 2^{2(i - 1)}$ elements to $S_i$ and removes at most $d \cdot 2^{2(i - 1)}$ elements from $B_i$.
Hence, after 4 executions of $\textsc{Resolve}(i - 1)$, $|S_i| \leq 4d \cdot 2^{2(i - 1)} = d \cdot 2^{2i}$ and $|B_i| \geq d \cdot 2^{2i + 1} - 4d \cdot 2^{2(i - 1)} = d \cdot 2^{2i}$.
A similar argument is made for the preconditions on $B_{i+1}$ and $S_{i+1}$.
\end{proof}

After each operation, we must perform a $\textsc{Resolve}(0)$ to ensure that the preconditions ($S_0$ is empty and $B_0$ is non-empty) are met for the next operation.
Thus, each $\textsc{Op}_{k}$ depends on $\res{0}{k-1}$, i.e., $\res{0}{k-1} \rightarrow \textsc{Op}_{k} \rightarrow \res{0}{k} \Rightarrow \res{0}{k-1} \rightarrow \res{0}{k}$.
Furthermore, recall that $\textsc{Resolve}(i)$ can modify $B_i$, $S_i$, $B_{i+1}$, and $S_{i+1}$, therefore, concurrent access to these arrays need to be avoided during parallel executions of \textsc{Resolve}.
In other words, no two consecutive levels can execute \textsc{Resolve} concurrently, i.e., $\res{i}{k} \rightarrow \res{i-1}{4k+1}$.

\section{Analysis}
\label{sec:analysis}
\subsection{PRAM Analysis}
\label{sec:pram}
\begin{lemma}
\label{lemma:pram-resolve-time}
Let $D(R_i)$ be the depth and $W(R_i)$ be the work of $\textsc{Resolve}(i)$.
For all $i \geq 0$ and $d \geq 1$, $D(R_i) = \BigO{\max(1, i + \log{d})}$ and $W(R_i) = \BigO{d \cdot 2^{2i}}$.
\end{lemma}
\begin{proof}
The \textsc{Resolve} operation relies on performing \textsc{Merge}, \textsc{Select}, and \textsc{DeleteDuplicates} on elements in levels $i$ and $i+1$.
\textsc{DeleteDuplicates} involves identifying and deleting duplicate entries and compressing the remaining elements into contiguous space, which can be accomplished via a parallel scan and prefix sum.
Hence, \textsc{Merge}, \textsc{Select}, and \textsc{DeleteDuplicates} on $n$ total elements can be performed with $D(n) = \BigO{\log{n}}$ depth and $W(n) = \BigO{n}$ work~\cite{pram}.
Therefore, $D(R_i) = \BigO{\log{(|B_{i+1}| + |S_{i+1}|)}} = \BigO{i + \log{d}}$ and $W(R_i) = \BigO{|B_{i+1}| + |S_{i+1}|} = \BigO{d \cdot 2^{2i}}$.
\end{proof}

\begin{lemma}
\label{lemma:pram-startend}
Let $c > 0$ be some constant, $T(R_0)$ be the time it takes to execute $\op{k}$ and $\textsc{Resolve}(0)$, and $T(R_i)$ be the time it takes to execute $\textsc{Resolve}(i)$.
For any $T(R_i) \leq c d \cdot 2^{2i}$, $\res{i}{k}$ completes execution before time 
$$\Paren{5k \cdot 4^i - 5 + \frac{1}{3}(4^i-1)} \cdot T(R_0) + T(R_i)$$
\end{lemma}
\begin{proof}
We know from the dependencies that for $k \pmod 4 \not\equiv 1$, $\res{i}{k}$ cannot start until $\res{i-1}{4k}$ finishes execution; and for $k \pmod 4 \equiv 1$, $\res{i}{k}$ cannot start until $\res{i-1}{4k}$ and $\res{i+1}{(k-1)/4}$ finishes execution.
Using induction, $\res{i-1}{4k}$ completes execution before time 
\begin{gather*}
\Paren{5(4k) \cdot 4^{i-1} - 5 + \frac{1}{3}(4^{i-1} - 1)} \cdot T(R_0) + T(R_{i-1}) \\
\leq \Paren{5k \cdot 4^i - 5 + \frac{1}{3}(4^{i-1} - 1)} \cdot cd + \Paren{cd \cdot 2^{2(i-1)}} \\
= \Paren{5k \cdot 4^i - 5 + \frac{1}{3}(4^{i-1} - 1) + 2^{2(i-1)}} \cdot cd \\
= \Paren{5k \cdot 4^i - 5 + \frac{1}{3}(4^i - 1)} \cdot T(R_0)
\end{gather*}
and $\res{i+1}{(k-1)/4}$ completes execution before time
\begin{gather*}
\Paren{5\Paren{(k-1)/4} \cdot 4^{i+1} - 5 + \frac{1}{3}(4^{i+1}-1)} \cdot T(R_0) + T(R_{i+1}) \\
\leq \Paren{5k \cdot 4^i - 5 \cdot 4 ^i - 5 + \frac{1}{3}(4^{i+1}-1)} \cdot cd + \Paren{cd \cdot 2^{2(i+1)}} \\
= \Paren{5k \cdot 4^i - 5 \cdot 4 ^i - 5 + \frac{1}{3}(4^{i+1}-1) + 2^{2(i+1)}} \cdot cd \\
= \Paren{5k \cdot 4^i - 5 + \frac{1}{3}(4^i - 1)} \cdot T(R_0)
\end{gather*}
Therefore, $\res{i}{k}$ completes execution before time $\Paren{5k \cdot 4^i - 5 + \frac{1}{3}(4^i - 1)} \cdot T(R_0) + T(R_i)$.
\end{proof}

\begin{theorem}
\label{thm:pram-result}
$\textsc{extractMin}$, $\textsc{update}$, $\textsc{delete}$, and $\textsc{bulkUpdate}$ on up to $d$ elements, has an amortized parallel depth of $O(1 + \log{d})$ and $\BigO{d\log_4{\frac{n}{d}}}$ work.
\end{theorem}
\begin{proof}
From Lemma~\ref{lemma:pram-startend}, $N$ operations complete execution before time $(5N - 4) \cdot T(R_0)$.
On a machine with infinite processors, $T(R_0) = D(R_0) = O(1 + \log{d})$, and the depth is $O(N\log{d})$ or an amortized $O(1 + \log{d})$ per operation.
Since the \struct has a total of $\Ceil{\log_4{\frac{n}{d}}} + 1$ levels, $O(\log_4{\frac{n}{d}})$ levels may be active in each step.
Hence, for a single processor, $T(R_0) = W(R_0) = O(d)$, and the \struct performs a total of $\BigO{Nd\log_4{\frac{n}{d}}}$ work or an amortized $\BigO{d\log_4{\frac{n}{d}}}$ per operation. 
\end{proof}

\subsection{I/O Analysis}
\label{sec:pem}
To optimize the I/O performance of the \struct, we set $d = O(M)$, so that $S_0$, $B_0$, and an additional buffer of size $d$ can always be maintained in a single processor's internal memory space.
For a single processor, $\textsc{Resolve}(i)$ can be performed using scans of contiguous memory, hence, $\textsc{Resolve}(i)$ performs $\BigO{\frac{d \cdot 2^{2i}}{B}}$ I/Os.

\begin{theorem}
\label{thm:pem-result}
For $1 \leq p \leq \Ceil{\log_4{\frac{n}{d}}} + 1$,
the \struct can perform \textsc{ExtractMin}, \textsc{Delete} and \textsc{Update} using 
$$\BigO{\frac{\log_4{n/d}}{pB} + \frac{1}{B}}$$
amortized parallel I/Os.
\end{theorem}
\begin{proof}
Let $p = \Ceil{\log_4{\frac{n}{d}}} + 1$.
We assign processor $p_i$ to level $i$ of the \struct.
In particular, processor $p_0$ is assigned to the first level of the heap and it always maintains $S_0$, $B_0$, and the auxilliary buffer of size $d$ in internal memory.
This additional buffer of size $d$ is used as an intermediate storage space for elements that will be merged into $S_1$ during the next call to $\textsc{Resolve}(0)$.
Hence, the first level of the heap is able to process $\Theta(d)$ \textsc{ExtractMin}, \textsc{Delete} and \textsc{Update} operations in internal memory before calling $\textsc{Resolve}(0)$.
We apply the resolution schedule described in Section~\ref{sec:pram}, 
where $T(R_0) \leq \frac{cd}{B}$ is the number of parallel I/Os performed by $\textsc{Resolve}(0)$.
As we perform $\Theta(d)$ operations for each $\textsc{Resolve}(0)$, $N$ operations can be performed using $N/d$ calls to $\textsc{Resolve}(0)$.
Thus, $\res{0}{N/d}$ finishes execution before time $\Paren{\frac{5N}{d} - 4} \cdot T(R_0)$.
Therefore, performing $N$ operations takes $\BigO{\frac{N}{d} \cdot \frac{d}{B}} = \BigO{\frac{N}{B}}$ parallel I/Os; or $\BigO{\frac{1}{B}}$ per operation.

Let $1 \leq p < \Ceil{\log_4{\frac{n}{d}}} + 1$.
Similar to the first case, we assign processor $p_0$ to the first level of the heap.
The remaining levels are divided equally across the remaining processors, so that each processor (except $p_0$) maintains $\BigO{\frac{\log_4{n/d}}{p}}$ levels.
In the resolution schedule, 
each processor performs all of the work associated with the levels it is assigned.
Therefore, performing $N$ operations takes $\BigO{\frac{N}{d} \cdot \frac{d}{B} \cdot \frac{\log_4{n/d}}{p}} = \BigO{\frac{N}{pB} \cdot \log_4{\frac{n}{d}}}$ parallel I/Os; or $\BigO{\frac{\log_4{n/d}}{pB}}$ per operation.
\end{proof}

\begin{theorem}
For $1 \leq p \leq \Ceil{\log_4{\frac{n}{d}}} + 1$,
the \struct can perform $\textsc{BulkUpdate}(U)$ on up to $d$ elements using
$$\BigO{\frac{d}{pB} \cdot \log_4{n/d} + \frac{d}{B}}$$
amortized parallel I/Os.
\end{theorem}
\begin{proof}
From Theorem~\ref{thm:pem-result}, it follows that $N$ $\textsc{BulkUpdate}(U)$ and $\textsc{Resolve(0)}$ operations finishes execution before time $\Paren{5N - 4} \cdot T(R_0)$.
Therefore, for $p = \Ceil{\log_4{\frac{n}{d}}} + 1$, performing $N$ operations takes $\BigO{\frac{Nd}{B}}$ parallel I/Os; or $\BigO{\frac{d}{B}}$ per operation.
And for $1 \leq p < \Ceil{\log_4{\frac{n}{d}}} + 1$, performing $N$ operations takes $\BigO{\frac{Nd}{B} \cdot \frac{\log_4{n/d}}{p}} = \BigO{\frac{Nd}{pB} \cdot \log_4{n/d}}$ parallel I/Os; or $\BigO{\frac{d}{pB} \cdot \log_4{n/d}}$ per operation.
\end{proof}

\section{Experiments}
\label{sec:experiments}
\subsection{GPU Implementation Details}
\label{sec:implementation}
From the analysis performed in Section~\ref{sec:pram}, the \struct is able to support a large number of threads in the PRAM model.
However, the parallel I/O performance (in Section~\ref{sec:pem}) relies on a relatively small number of processors, $1 \leq p \leq \Ceil{\log_4{\frac{n}{d}}} + 1$ (i.e., at most 1 processor per level of the \struct), due to a single processor needing to maintain the first level of the heap in internal memory.
This restriction of processors in the PEM model can be an issue in traditional many-core architectures (e.g., multi-core CPU systems), however, the thread and memory hierarchy of GPUs allows us to harness both the parallelism shown in the PRAM analysis and I/O efficiency shown in the PEM analysis.
As all I/Os performed in the \struct are scans of contiguous memory locations, we are able to schedule warps such that warps belonging to the same thread block access contiguous blocks of $w$ elements.
Hence, we map each thread block consisting of $tw$ threads (or $t$ warps) to a single PEM processor (where shared memory is used as the internal memory space) and use a block size of $B = tw$.

At the start of execution, all thread blocks are launched onto the GPU and execute $\textsc{Empty}(S_i)$ and $\textsc{Fill}(B_i)$ as needed throughout the course of the program.
Because CUDA C/C++ only provides hardware synchronization primitives between threads within a thread block (i.e. intra-block synchronization), synchronization between threads across thread blocks (i.e., inter-block synchronization) can only be performed via software implemented synchronizations.
Past implementations of software synchronizations between thread blocks~\cite{garlandlock,interblock,lockbased} show that variations of busy-wait (i.e., spin locks) can be used to communicate synchronization information (e.g. the number of available resources or the number of thread blocks that have reached the synchronization point).
We use a similar approach, where each thread block (i.e., level of the \struct) has a designated global memory location that is used to signal the particular thread block to execute $\textsc{Empty}(S_i)$ and/or $\textsc{Fill}(B_i)$.
Hence, a single thread per thread block can be used to continuously check this memory location until it has been set to a particular value.
To avoid deadlocks, this approach requires that all thread blocks are able to be concurrently scheduled onto the GPU.
A simple way to ensure this is to never launch more thread blocks than there are SMs.

As shown in the psuedocode of $\textsc{Resolve}(i)$ (Algorithm~\ref{alg:resolve}), implementing $\textsc{Empty}(S_i)$ and $\textsc{Fill}(B_i)$ requires using parallel subroutines: \textsc{Merge}, \textsc{PrefixSums}, and \textsc{Select}.
We use the implementation of \textsc{Merge} provided in the Thrust library~\cite{thrust} and the implementation of \textsc{PrefixSums} provided in the CUB library~\cite{cub}.
We could not find a high-performance GPU implementation of \textsc{Select}, hence, we instead use the implementation of \textsc{RadixSort} provided in the CUB library (which trivially allows us to find the $k$-th smallest priority in an array after it is sorted).
These parallel subroutines are called from each thread block using CUDA dynamic parallelism.

The \struct is used to solve the SSSP problem using a parallel variant of Dijkstra's algorithm, denoted \sssp.
Given a graph with $n$ total vertices and maximum degree $d$, we perform $n$ rounds where in each round: 
(1) the minimum distance vertex, denoted $u$, is extracted from the \struct; 
(2) all outgoing edges $(u, v)$ are relaxed in parallel; and 
(3) all edges $(u, v)$ that resulted in a shorter distance to $v$ are inserted into the \struct via a \textsc{BulkUpdate} operation.
This algorithm can be implemented using parallel scans and \textsc{PrefixSums} (on a maximum of $d$ elements).
To reduce the number of \textsc{BulkUpdate} operations, we set the maximum update batch size to be equal to $d$.

\subsection{Methodology}
\label{sec:methodology}
We experimentally compare the performance of \sssp to the state-of-the-art SSSP GPU implementations, Gunrock~\cite{gunrock} and Asynchronous Dynamic Delta-Stepping (ADDS)\footnote{The ADDS library does not support compute capability 5.2}~\cite{adds}, both of which are GPU implementations of the Delta-stepping algorithm.
Meyer et al.~\cite{delta-stepping} proved that for an arbitrary graph with random edge weights, the performance of the Delta-stepping algorithm is a function of the maximum degree $d$ and maximum shortest path length $L$ of the input graph.
Moreover, past experimental work~\cite{busato,owens-sssp} showed that the performance of previous implementations of Delta-stepping on GPUs degrade on sufficiently dense graphs with large diameters (i.e., the number of edges in the maximum shortest path).
In this work, we demonstrate that using the parallel cache-efficient \struct in a parallel variant of Dijkstra's algorithm provides a suitable implementation for solving SSSP on GPUs for such graphs.
Therefore, our experiments are conducted on graphs with sufficiently large degrees and diameter.

We generate 5 random directed acyclic graphs (DAGs) containing $n = $ 30,000 vertices with diameter and maximum shortest path distance of $L = n-1$.
For 1-indexed vertices (i.e., vertices are identified via integers 1, 2, $\ldots$, n), we generate the random graphs with $m \geq n-1$ edges in two stages.
In the first stage, the shortest paths (and diameter) are created such that for each $i \in \{1, 2, \ldots, n-1\}$, edge $(i, i+1)$ with weight 1 is inserted into the graph.
Afterwards, the remaining $(m-n+1)$ edges are generated randomly with the following constraints: 
edges are distributed uniformly across the vertices $i \in \{1, 2, \ldots, n-1\}$, such that vertex $i$ has a maximum degree of $(n-i)$; and 
if edge $(u, v)$ is generated, then its weight is set to $(2 \cdot (u-v))$ to ensure that the shortest paths (and diameter) of the graph remains unchanged.
Using these generated random graphs, the depth and work of the Delta-stepping algorithm becomes $O(dn\log{n} + \log^2{n})$ and $O(n + m + dn\log{n})$, respectively, in the average case.
In comparison, \sssp (using the \struct) has a parallel depth of $O(n(1 + \log{d}))$ and $\BigO{n + m + dn\log{\frac{n}{d}}}$ work, when running on the generated random graphs.

All code is compiled using CUDA C/C++ 11 and experiments are performed on 2 Nvidia GPUs: an RTX 2080 Ti (compute capability 7.5), containing 4,352 physical processors distributed across 68 SMs, 11 GB of global memory, and 96 KiB of unified L1 cache and shared memory per SM; and a Quadro M4000 (compute capability 5.2), containing 1,664 physical processors distributed across 13 SMs, 8 GB of global memory, and 96 KiB of shared memory per SM.
All runtime experiments are conducted on each of the generated input graphs, where for each graph, 10 trials are conducted.
The average runtime across all trials (for all input graphs) are reported.

\subsection{Runtime Results}
\label{sec:results}
\begin{figure*}[tb]
  \centering
  \includegraphics[width=0.75\textwidth]{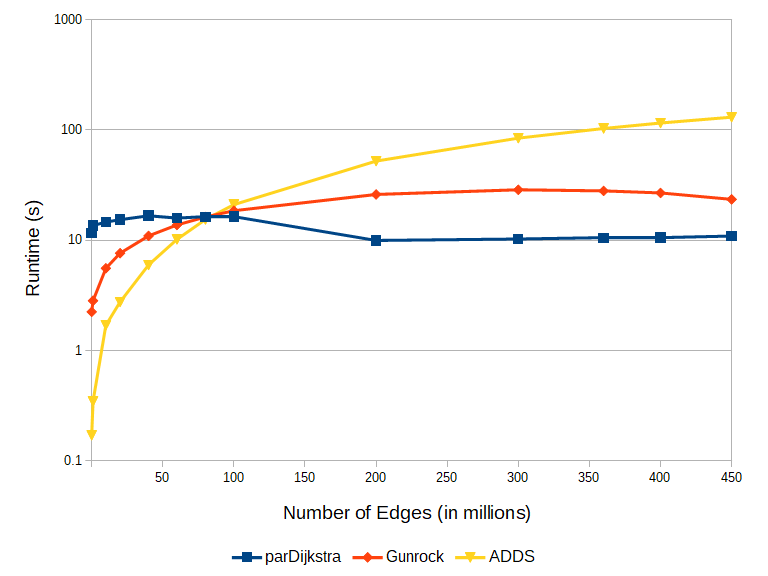}
  \caption{Average runtime (in seconds) on the generated random DAGs with $n = 30$ thousand vertices and diameter $L = n$ on an NVIDIA RTX 2080 Ti.}
  \label{fig:2080ti-runtime}
\end{figure*}

\begin{figure*}[tb]
  \centering
  \includegraphics[width=0.75\textwidth]{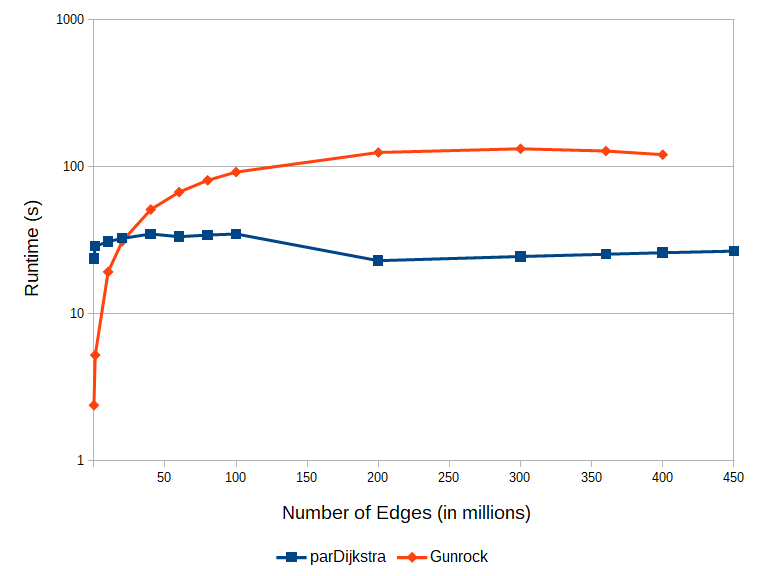}
  \caption{Average runtime (in seconds) on the generated random DAGs with $n = 30$ thousand vertices and diameter $L = n$ on an NVIDIA Quadro M4000.}
  \label{fig:m4000-runtime}
\end{figure*}

Figure~\ref{fig:2080ti-runtime} and Figure~\ref{fig:m4000-runtime} plots the average runtime of each of the SSSP algorithms across all generated random DAGs, compared to the number of edges in the input graph, on an NVIDIA RTX 2080 Ti and Quadro M4000, repsectively.
Since edges are distributed uniformly in the generated random DAGs, as the number of edges increase, the maximum (and average) degree of the input graph also increases.
Additionally, as the impact of I/O efficiency on overall runtime is pronounced on large input sizes (i.e., a large number of edges), we expect the \sssp using the I/O efficient \struct to perform well on these inputs.

Results show that on the RTX 2080 Ti, \sssp is faster than Gunrock and ADDS once the number of edges in the input graph exceeds 80 million edges and 100 million edges, respectively.
And on the Quadro M4000, \sssp is faster than Gunrock once the the number of edges in the input graph exceeds 40 million edges.
Furthermore, we find that while the ADDS implementation is faster than Gunrock for less than 80 million edges, its performance degrades significantly compared to Gunrock for graphs with a larger number of edges.
On the RTX 2080 Ti, we observe a peak speedup of 2.8 compared to Gunrock and a peak speedup of 12 compared to ADDS, on the graphs with 300 million edges and 450 million edges, respectively.
The peak speedup of \sssp compared to Gunrock on the Quadro M4000 is 5.4, occurring at 200 million edges.  
We note that on the Quadro M4000, Gunrock was unable to run on the graphs with 450 million edges, due to an out-of-memory error.

\section{Conclusion}
\label{sec:conclusion}
In this paper, we have presented the parallel bucket heap, denoted \struct, a parallel variant of the cache-oblivious bucket heap~\cite{brodal-bucketheap} and buffer heap~\cite{chowdhury-buffer}.
The parallel bucket heap supports standard priority queue operations: \textsc{Update}, \textsc{Delete}, and \textsc{ExtractMin}, as well as \textsc{BulkUpdate} of up to $d$ elements.
For a maximum of $n$ elements in the \struct, all operations can be performed with an amortized depth of $O(1 + \log{d})$ and $O(d\log_4{\frac{n}{d}})$ work in the CREW PRAM model.
To optimize for I/O efficiency, $d$ is bounded by the internal memory size of a processor (i.e., $d = O(M)$), resulting in $\BigO{\frac{\log_4{n/d}}{pB} + \frac{1}{B}}$ amortized parallel I/Os per \textsc{Update}, \textsc{Delete}, or \textsc{ExtractMin} operation; and $\BigO{\frac{d}{pB} \cdot \log_4{\frac{n}{d}} + \frac{d}{B}}$ amortized parallel I/Os per \textsc{BulkUpdate} operation, in the PEM model.

We implement the \struct on the GPU using CUDA C/C++ and use it in a parallel variant of Dijkstra's algorithm for solving the SSSP problem, denoted \sssp.
Experimental results show that on an Nvidia RTX 2080 Ti and Quadro M4000, the \sssp outperforms the current state-of-the-art SSSP GPU implementations, Gunrock~\cite{gunrock} and ADDS~\cite{adds}, on our generated random DAGs with $n = 30,000$ vertices and diameter $n-1$.
On the RTX 2080 Ti, we observe a peak speed up of 2.8 and 12 compared to Gunrock and ADDS, respectively; and on the Quadro M4000, \sssp provides a peak speed up of 5.4 compared to Gunrock.
This work highlights the unique architecture of GPUs and how the thread and memory hierarchy can be leveraged to obtain both I/O efficiency per thread block and a high degree of parallelism.

\bibliographystyle{plain}
\bibliography{mybib}

\end{document}